\documentclass[aps,prb,amssymb,showpacs,twocolumn,amsfonts]{revtex4}
\usepackage{tabularx,graphicx,float}

\usepackage{graphicx,amsmath}
\usepackage{amssymb}
\usepackage{epsfig}

\date{\today}

\begin{document}

\bibliographystyle{apsrev}

\author{R. Rold\'{a}n$^{1,2}$, Shan-Wen Tsai$^3$ and M. P. L\'opez-Sancho$^2$}
\affiliation{$^1$Laboratoire de Physique des Solides, Univ. Paris-Sud, CNRS, UMR 8502, F-91405 Orsay Cedex, France.\\
$^2$Instituto de Ciencia de Materiales de Madrid, CSIC,
Cantoblanco, E-28049 Madrid, Spain\\
$^3$Department of Physics and Astronomy, University of California, Riverside, CA 92521, USA.
}

\title{Renormalization group approach to anisotropic superconductivity} 
\begin{abstract}

The superconducting instability of the Fermi liquid state is investigated by considering anisotropic electron-boson
couplings. Both  electron-electron interactions and anisotropic electron-boson couplings
are treated with a renormalization-group method that takes into account retardation effects.
Considering a non-interacting  circular Fermi surface, 
we find analytical solutions for the flow equations and derive a set of generalized Eliashberg 
equations. Electron-boson couplings with different momentum
dependences are studied, and we find  superconducting instabilities of the
metallic state with competition between order parameters of different symmetries.
Numerical solutions for some couplings are given to illustrate
the frequency dependence of the vertices at different coupling regimes.

\end{abstract}

\maketitle

\section{Introduction}

Many-body interactions have important effects on the electronic properties of
correlated materials and are subject of great attention. 
The main role played by the electron-phonon interaction in the conventional 
superconductivity has enhanced the interest in electron-boson coupling.
In the formation of Cooper electron pairs in conventional superconductors,
the attractive pairing interaction is mediated by phonons, as established by the Bardeen
Cooper Schriffer (BCS) theory of superconductivity. The Eliashberg theory
\cite{E60,C90} provides the appropriate equations to obtain the superconducting temperature at which pairing
occurs, as well as the energy gap created in the electronic density of states (DOS). The 
Eliashberg equations describe an effective electron-electron interaction due to the 
exchange of any form of bosons. The Eliashberg function $\alpha^2F(\omega)$, which
gives the spectral electron-phonon density, and the Coulomb pseudo potential $\mu^*$
(Anderson-Morel potential\cite{MA62}), which characterize the pairing interaction, 
are input parameters of the Eliashberg equations.

Direct evidence that the bosons mediating the attractive interaction, in conventional
superconductors, are phonons was provided by tunneling experiments \cite{McMR65,McMR69,Scal69}. Following McMillan and Rowell work\cite{McM68,McMR69}, by inversion of tunneling data, a unique value of $\alpha^2F(\omega)$, as well of $\mu^*$, is provided by the 
structure in the electron tunneling current measured as a function of the applied voltage,
the $dI/dV$ characteristic. For BCS superconductors,  the gap created in the DOS when pairing
takes place is  isotropic, of $s$-wave   symmetry. However, anisotropic electron-phonon coupling has been reported, later on,
in some borocarbide material superconductors as YNi$_2$B$_2$C and LuNi$_2$C
\cite{MMEHS01, MS01}.  Furthermore the superconducting energy gap of  MgB$_2$
was found to vary on the Fermi surface (FS) due to the momentum dependence of the electron-phonon interaction.
The superconducting properties of  this material, which presents the highest superconducting 
transition temperature ($T_c=39$K) among the binary compounds \cite{BY01,NNMZA01}, are explained
by the fully anisotropic Eliashberg theory\cite{AM82,C90}, by including the momentum dependence
of the electron-phonon coupling, combined with density functional calculations \cite{CRSCL02}.

The anisotropic Eliashberg theory has also been applied to materials with  other pairing
symmetries\cite{MSV88,BZ06} such as the copper oxides which present a $d_{x^2-y^2}$-wave order parameter. In this case,
besides the phonons, other kind of bosons have been considered as, for example, spin fluctuations \cite{M92}.
In the copper oxides high-$T_c$ superconductors, there is currently not a consensus about the nature 
of the pairing interaction. The  $d$-wave symmetry gap  experimentally measured, seems
to favor a purely electronic pairing interaction. However a kink, i.e., a change of the slope in the quasiparticle energy dispersion, has been
reported by ARPES experiments \cite{L01} at an energy of about 50meV. This kink could indicate a 
renormalization effect similar to that appearing in conventional superconductivity. In the cuprates, at this energy scale, there are multiple excitations,
such as phonons and spin fluctuations, which could be responsible for the renormalization. At higher
energy, 350 meV, other kink has been reported related with bosonic excitations\cite{V06}.
Both, phonons\cite{L01} and spin fluctuations\cite{B03}, have been proposed 
but the nature of the bosonic mode remains under debate. Therefore strong electron-electron 
correlations as well as strong electron-boson coupling have to be considered in order to
analyze the experimental data\cite{HFL07}.

Experiments carried out  by different techniques  on hole-doped cuprates, with
oxygen isotope substitution ($^{16}$O-$^{18}$O) \cite{GSZGTLL04,LFMS06,KSCP07},
have claimed that phonons play an important role in cuprates.
Recent data have provided evidence that the electron-phonon interactions
are responsible for the origin of the nodal kink \cite{IDSM08}, confirming earlier
ARPES results \cite {DCSN04} which had indicated an anisotropic electron-phonon
interaction in cuprates in both the normal and superconducting states.
However, the mechanism of pairing in high temperature superconductors
is still under discussion.

We focus here in the study of the superconducting instability with pairing mediated
by anisotropic electron-boson coupling. The most general case of anisotropic coupling
of electrons to bosonic modes is considered, with a combination of many symmetry channels.
We follow the asymptotically exact renormalization group (RG) approach developed 
in ref.\cite{Shan05}. This extension of the RG
treats electron-electron and electron-phonon interaction on an equal footing.
Analytical and numerical solutions of the flow equations for the BCS vertices are obtained.
Generalized Eliashberg equations and the corresponding MacMillan-Rowell  
and Allen-Dynes expressions for $T_c$ are derived from the RG flow equations.
The Migdal theorem\cite{Migdal58} assumed in the Eliashberg theory, is understood in terms
of a large-$N$ expansion in the RG approach\cite{Shan05}. We include both, retardation effects and the presence 
of multiple energy scales in the problem. Retardation effects in
 the one-dimensional (1D) Holstein-Hubbard model at half-filling have been found to be
important near the transition between spin density wave (SDW) and 
charge density wave (CDW) states
\cite{TTCC07}. The classical and quantum aspects of fermion driven lattice 
instabilities have been analyzed using an extension of the RG method 
similar to the one followed in the present work \cite{BB07}. The calculation of the gap and
the structure of the phase diagram of 1D
tight-binding models, such as molecular crystal and Su-Schrieffer-Heeger models,
have been analyzed at the one-loop level\cite{BB07}. Furthermore, RG methods have been recently used to study the Cooper instability in graphene\cite{H08,G08}.

The paper is organized as follows. In section II the RG analysis is developed.
In Section III the generalized Eliashberg equations are discussed . Section IV addresses results
obtained for electron-boson couplings with different angular dependences.
Finally in Section V the work is summarized and some conclusions are given.

\section{Renormalization-group analysis.}

The RG method for interacting electrons\cite{S94} is adopted here to analyze the
instabilities of a two-dimensional (2D) Landau Fermi liquid. We start from a given geometry of the Fermi surface of the
non-interacting system and we treat the microscopic interactions within one-loop
RG, which captures the essential physics of the problem. In this way,  assumption
of a predetermined  order parameter is not needed. Since we are interested in the superconducting instability, both electron-electron
and electron-boson interactions have to be considered. Therefore we follow the asymptotically exact RG scheme 
extended to include interacting fermions coupled to bosonic modes\cite{Shan05}. 

In the Wilson-like RG theory\cite{S94}, at a first step, the modes of the momentum space above a cutoff
$\Lambda$ are integrated out. The remaining phase space in a 2D system is a ring
of radius $k_F$ and width of 2$\Lambda$ around the Fermi energy $E_F$. 
In the second step, the equations are solved by a large-$N$ method, with $N=E_F/\Lambda$,
$N$ being the number of patches at the Fermi surface. Here a 2D square lattice is considered, at low fillings, so that the FS has an almost
circular shape. The interactions are parametrized
by the on-site Coulomb repulsion $u_0$, the electron-boson
coupling $g$, and the Einstein frequency $\omega_E$ of the bosons.
In this case the electron-boson constant is $\lambda = 2N(0)g^2/\omega_E$, where $N(0)$ 
is the density of states at the Fermi level.

The RG equations are derived in the path-integral representation\cite{Shan05}. In order to obtain a 
combination of many symmetry channels, the most general anisotropic electron-boson 
coupling $g(i,j)$, is considered. The electron-boson coupling can be integrated out exactly,
leading to an electron-electron effective problem with retarded interactions.
The retarded electron-electron interaction has the form:
\begin{eqnarray}\label{RetadEq}
\widetilde{u}(4,3,2,1) = u(4,3,2,1) - 2g(1,3)g(2,4)D(1-3)
\end{eqnarray}
where the phonon propagator is
\begin{eqnarray}\label{PhopEq}
D(q) = \frac{\omega_{\mathbf q}}{\omega^2 + \omega_{\mathbf q}^2}.
\end{eqnarray}
We adopt  the notation $q = (\omega,\mathbf{q})$, $1=(\omega_1,\mathbf{k}_1)$, and so forth.
Spin indices are omitted in this notation.
Particles 1 and 2 are incoming and scatter into 3 and 4, respectively. 
We will focus on processes involving particles with opposite spins.
The processes involving particles with same spins can be obtained from these due to
SU(2) symmetry\cite{ZS00}.
For our almost  circular FS, the effective retarded electron-electron interaction
$\widetilde{u}$ gives  two types of scattering in the RG: the forward scattering,
with $k_1=k_3$, $k_2=k_4$ which, as in the case of pure electron-electron interaction,
does not get renormalized but contributes to the electron self-energy, and the scattering in the Cooper channel. While the forward channel does not 
flow under the RG, the BCS vertex ($k_1=-k_2$, $k_3=-k_4$) flows at one-loop in the RG
equation:
\begin{eqnarray}\label{FlowEq}
\frac{d}{d\ell}\widetilde{u}(1,3,\ell)\! = \! - \!\! \int_{-\infty}^{\infty}
\!\!\! \frac{d\omega}{\pi} \!\! \int_{0}^{2\pi} \!\!\! \frac{d\theta}{2\pi}
\frac{\Lambda_{\ell} \widetilde{u}(1,\omega,\theta,\ell)
 \widetilde{u}(\omega,\theta,3,\ell)}{\Lambda_{\ell}^2 +
  Z_{\ell}^2(\omega,\theta)\omega^2}
\end{eqnarray}
 where  $1 \equiv (\omega_1,\theta_1)$ and $3 \equiv (\omega_3,\theta_3)$,
 and $\ell$ represents the RG scale $\ell=\ln(\Lambda_0/\Lambda)$, with 
 $\Lambda_0$ the initial bandwidth. Eq. (\ref{FlowEq}) is obtained from
 the cutoff independent condition imposed to the interaction vertex
 \begin{equation}
 \frac{d}{d\ell}\widetilde{u}(4,3,2,1)=0.
 \end{equation}
The vertices are labeled  according to the angle around
the FS $\theta_i$, since the dependence on the radial part of the momentum is
irrelevant here\cite{S94}. However, the dependence on the modulus $|{\bf q}|$ is important at larger fillings\cite{GMV04,NTS08} where, in addition, electron-electron interactions may lead to renormalization of the FS shape\cite{R06b}. In order to solve the differential equation (\ref{FlowEq}),
the pairing potential is decomposed in terms of the irreducible representation of 
the space group of the underlying lattice. For the case of a 2D square
lattice, the space group is $D_4$. The irreducible
representations of the $D_4$ space group contains four one-dimensional (singlets):
$A_1$ and $A_2$ for conventional and unconventional
$s$-wave channels, $B_1$ for $d_{x^2-y^2}$-channel and $B_2$
for $d_{xy}$-channel; and one two-dimensional (triplet):
$E$, corresponding to $p$-wave symmetry, with degenerate eigenvalues
for the two channels $p_x$ and $p_y$. The corresponding basis functions are chosen to be:
\begin{eqnarray}\label{Fi}
  A_1&&  \phi^1_n(\theta)  = C_n\cos[4(n-1)\theta] \nonumber\\
  B_1&&  \phi^2_n(\theta) = C_0\cos[(4(n-1)+2)\theta]\nonumber \\
  B_2&&  \phi^3_n(\theta) = C_0\sin[(4(n-1)+2)\theta] \nonumber\\
  A_2&&  \phi^4_n(\theta) = C_0\sin[4n\theta] \nonumber\\
  E  &&\left\{ \begin{array}{cc}
      \phi^5_n(\theta)=&C_0\sin[(2n-1)\theta]\nonumber\\
      \phi^6_n(\theta)=&C_0\cos[(2n-1)\theta]\\
    \end{array}\right. \\
\end{eqnarray}
with $n$ ranging from 1 to $\infty$, and the normalization factors $C_n=1/\sqrt{\pi(1+\delta_{n1})}$.
Therefore we can express the pairing potential 
in the more general form
\begin{eqnarray}\label{Ppeq}
\widetilde{u}(i,j,\ell) = \sum_{\gamma}
\sum_{m,n} u_{mn}^{\gamma}(\omega_i,\omega_j,l)
f_{mn}^{\gamma}(\theta_i,\theta_j),
\end{eqnarray}
where $\gamma$ labels the representation of the group and
\begin{eqnarray}
f_{mn}^{\gamma}(\theta_i,\theta_j) =
\phi_m^{\gamma}(\theta_i)\phi_n^{\gamma}(\theta_j),
\end{eqnarray}
where  $\phi_m^{\gamma}(\theta)$ are the corresponding basis functions given in Eq. (\ref{Fi}).
The normalized basis functions $\phi_m^{\gamma}(\theta)$ are  orthogonal 
\begin{equation}
\int_0^{2\pi}\phi_m^{\gamma}(\theta)\phi_n^{\gamma'}(\theta)d\theta=
\delta_{mn}\delta^{\gamma\gamma'}
\end{equation}
and form a  complete  basis set. By discretizing the frequency integral into a sum, 
taking advantage of the orthogonality and completeness of the $\phi$-basis,
and restricting ourselves to frequencies below the cutoff, the two-dimensional
matrix $\widetilde{u}(i,j,\ell)$  expressed in Eq. (\ref{Ppeq}) can be transformed 
into a four-dimensional tensor 
$ \left[\widehat{u}_{nm}^{\gamma}\right]^{\alpha}_{\beta}(\ell)$. Therefore
we can rewrite the flow equation, Eq. (\ref{FlowEq}), as a \emph{tensor} equation:
\begin{eqnarray}
\frac{d}{d\ell} \!\!
\left[\widehat{u}_{nm}^{\gamma}\right]^{\alpha}_{\beta}\!(\ell)\! =
\!-\!\!\!\!\! \sum_{\delta=-N}^N \!\sum_{i,j=0}^M \!\!
\left[\widehat{u}_{ni}^{\gamma}\right]^{\alpha}_{\delta}\!(\ell)\!
\left[\widehat{K}_{ij}^{\gamma}\right]_{\delta}^{\delta}\!\!(\ell)
\!\left[\widehat{u}_{jm}^{\gamma}\right]^{\delta}_{\beta}\!(\ell),
\label{Tensor}
\end{eqnarray}
where we have imposed an upper limit to the harmonic index
$n\le M$, and to the frequency sum, $-N\le \delta \le N$. In Eq. (\ref{Tensor}) we have used
\begin{eqnarray}
\left[\widehat{u}_{nm}^{\gamma}\right]^{\alpha}_{\beta}(\ell)& =&
\int\!\!{d\theta_{\alpha}}\!\!\int\!\!{d\theta_{\beta}}\,
\widetilde{u}(\alpha,\beta,\ell)
\phi_n^{\gamma}(\theta_{\alpha})\phi_m^{\gamma}(\theta_{\beta}),\nonumber\\
 \left[\widehat{K}_{ij}^{\gamma}\right]_{\beta}^{\alpha}(\ell)&=&
 K_{ij}^{\gamma}(\omega_{\alpha},\ell)\delta_{\beta}^{\alpha},
\end{eqnarray}
with the kernel
\begin{equation}\label{kernel}
K_{ij}^{\gamma}(\omega,\ell)=\int_0^{2\pi}\!\!\!\!d\theta
\frac{\Lambda_{\ell}}{\Lambda_{\ell}^2+Z_{\ell}^2(\omega,\theta)\omega^2}
\phi_i^{\gamma}(\theta)\phi_j^{\gamma}(\theta) \ .
\end{equation}

Therefore Eq. (\ref{Tensor}) can be written in a compact form
as a $(2N+1)M\times(2N+1)M$ matrix equation. The flow equation have the form
\begin{equation}\label{Matfl}
\frac{d\mathbf{U}^{\gamma}}{d\ell} = - \mathbf{U}^{\gamma}{\cdot}
\mathbf{K}^{\gamma}{\cdot}\mathbf{U}^{\gamma}
\end{equation}
where the matrix indices $j$ are related to the frequency
$\alpha$, and harmonic $n$ indices, by  $\alpha={\rm IP}\left[(j-1)/M\right]+1$ and $n=j-(\alpha-1)M$, respectively, where IP denotes the integer part. From Eq. (\ref{Matfl}) we obtain one vertex flow
equation for each channel $\gamma$. Once we have the matrix form
 of the RG flow equations for the vertices,  it is possible
 to write the exact solution:
\begin{equation}\label{U(l)}
 \mathbf{U}^{\gamma}(\ell)=[\mathbf{1} + \mathbf{U}^{\gamma}(0) \cdot
 \mathbf{P}^{\gamma}(\ell)]^{-1} \cdot \mathbf{U}^{\gamma}(0) \ ,
 \end{equation}
 where $\mathbf{P}^{\gamma}(\ell)\equiv\int_0^\ell
 d\ell^{\prime}\mathbf{K}^{\gamma}(\ell^{\prime})$. There is an instability of the Fermi liquid state when
$\mathbf{U}^{\gamma}(\ell_c)\rightarrow\infty$ for $\ell=\ell_c$. This condition
is fulfilled when
\begin{equation}\label{det-eq}
\det\left[\mathbf{1} + \mathbf{U}^{\gamma}(0) \cdot \mathbf{P}^{\gamma}(l_c)
\right]=0,
 \end{equation}
 which is equivalent to solving the eigenvalue equation
\begin{equation}
\left[\mathbf{1} + \mathbf{U}^{\gamma}(0) \cdot
  \mathbf{P}^{\gamma}(\ell_c)\right] \cdot {\mathbf{v}^{\gamma}} = 0 \ .
  \label{eigenvector-equation}
\end{equation}
  The kernel $\mathbf{K}^{\gamma}(\ell)$ that appears in the vertex flow
  equations is given
  by Eq. (\ref{kernel}) and contains self-energy corrections. 
  The momentum-dependent imaginary
  part of the self-energy $\Sigma^{''}(\omega,{\mathbf k})$, with contributions 
  from all the components of the electron-boson
  coupling, is related to the quasiparticle-weight by
  $\Sigma_{\ell}^{''}(\omega,\theta)=[1-Z_{\ell}(\omega,\theta)]\omega$. 
  The imaginary  part of the self-energy  is
  renormalized by the bare forward vertices, and the
  flow equation for the quasiparticle-weight $Z_{\ell}(\omega,\theta)$ can be integrated to give:
\begin{widetext}
\begin{eqnarray}\label{QP-Weight-a}
 Z_{\ell}(\omega_{\alpha},\theta_{\alpha}) = 1+
 \frac{2}{\pi\omega_{\alpha}} 
 \int_{\Lambda_{\ell}}^{\Lambda_0}\!\!d\Lambda_{\ell^{\prime}}
 \!\int \!d\theta_{\beta} d\omega_{\beta} \frac{N(0)\tilde{u}_0(\alpha,\beta)
  Z_{\ell^{\prime}}\!(\omega_{\beta},\theta_{\beta})
  \omega_{\beta}}{\Lambda_{\ell^{\prime}}^2
  + Z_{\ell^{\prime}}^2(\omega_{\beta},\theta_{\beta})\omega_{\beta}^2},
\end{eqnarray}
\end{widetext}
where $\tilde{u}_0(\alpha,\beta)$ is the retarded electron-electron interaction 
\begin{eqnarray}
\tilde{u}_0(\alpha,\beta)=u_0-2g^2(\theta_{\alpha},\theta_{\beta})
 D(\omega_{\alpha}\!-\!\omega_{\beta}).
\end{eqnarray}
 The self-energy is angle-dependent
 and it appears as a term in the denominator of
 $\mathbf{K}^{\gamma}(\ell)$. Therefore, whereas only the $\gamma$-component
 $\mathbf{K}^{\gamma}(\ell)$ of
 the kernel contribute to a given channel $\gamma$, all $\gamma$-components of
 the microscopic electron-boson  coupling contribute to $\mathbf{K}^{\gamma}(\ell)$. The results of this section will be used in Sec. IV, where a numerical analysis of the RG flow equations, as well as the $\lambda$--$W_c$ phase diagram are studied, $W_c$ being the energy scale of the instability.

\section{Eliashberg equations.}

Eliashberg theory\cite{E60,C90} of superconductivity was originally
formulated to describe
phonon-mediated $s$-wave superconductors. By including momentum dependence
in the electron-boson interaction, the anisotropic Eliashberg formalism is obtained.
The Eliashberg function $\alpha^2F(\omega)$, which defines
the electron-boson  coupling $\lambda =
2\int_0^{\infty} \alpha^2F(\omega) d\omega/\omega$ carries, in the anisotropic formalism, 
momentum dependence. The fully anisotropic Eliashberg formalism has been successfully
used to study the superconducting properties of MgB$_2$. By solving numerically
the anisotropic Eliashberg equations, the superconducting transition temperature, the
momentum-dependent superconducting energy gap, and the momentum-dependent specific
heat have been obtained\cite{CRSCL02,CCL06}. Generalization of the
Eliashberg equations to $d$-wave superconductivity has as well been
done\cite{JSCBT96} and has been used to investigate the physics of cuprates \cite{SC08}. As it was shown in Ref. \cite{Shan05},
the  reduced Coulomb repulsion $\mu^*$ or Anderson-Morel potential emerges naturally from the RG
equations as $\mu^* = u_0/(1+\ell_E u_0)$, where
$\ell_E=\ln(\Lambda_0/\omega_E)$. The value of $T_c$, and the analytical forms of the McMillan\cite{McM68}
and Allen-Dynes\cite{LC71,AD75} expressions are obtained from the RG flow equations.

Here the instability conditions obtained in the preceding section
for the BCS vertex are mapped 
into the Eliashberg equations. The breakdown of the Fermi liquid state by the SC instability, due to the retarded
electron-electron interactions,  occurs at a finite temperature which turns out to be
the same critical temperature $T_c$ obtained from the Eliashberg theory. 
To investigate this instability, we solve the RG flow equations at finite temperature.
Since the SC instability is approached by decreasing the temperature, in the equations
such as (\ref{eigenvector-equation}), where integration
over $\Lambda$ is involved, this can be extended to zero. In this formalism, considering 
a finite value of the temperature $T$, the quasiparticle weight $Z_{\ell}(\omega,\theta)$ also presents
a $T$ dependence. Taken the appropriate integration limits in Eq. (\ref{QP-Weight-a})
we obtain a similar expression to Eq. (\ref{eigenvector-equation}), which gives the condition
for the appearance of the instability and allows us to obtain the expression for $T_c$.
The integrals over frequencies can be written as Matsubara sums ($
\int d\omega_{\beta}/(2\pi) \rightarrow T_c \sum_{\beta}$) 
in equations (\ref{eigenvector-equation})
and (\ref{QP-Weight-a}), leading to: 
\begin{widetext}
\begin{eqnarray}
v^{\gamma}(\omega_n,\theta)&=&\pi T_c \sum_{\omega_m}
  \frac{1
    }{|\omega_m|} \int_{\theta^{\prime}}
    \Gamma_{\theta,\theta^{\prime}}^{\gamma}(\omega_n-\omega_m)
    \frac{v^{\gamma}(\omega_m,\theta^{\prime})}{Z(\omega_m,
    \theta^{\prime}, T_c)},\nonumber \\
  Z(\omega_n,\theta, T_c) &=&1 + \pi T_c \sum_{\omega_m}
  \bar{\Gamma}_{\theta}(\omega_n-\omega_m).\label{eq:ge}
  \end{eqnarray}
  \end{widetext}
 These two expressions are the set of generalized Eliashberg equations at $T_c$, where
 \begin{eqnarray}
  \Gamma^{\gamma}_{\theta,\theta^{\prime}}(\omega_{\alpha}\!\!-\!\omega_{\beta})
  \!\!\!&\equiv&\!\!\! - \!\!\! \int_{\theta_{\alpha},\theta_{\beta}} \!\!\!\!\!\!\!\!\!\! N(0)
  \tilde{u}_0(\alpha,\beta)
  \eta^{\gamma}\!(\theta_{\alpha},\theta)
  \eta^{\gamma}(\theta_{\beta},\!\theta^{\prime})\nonumber\\
  \bar{\Gamma}_{\theta_{\alpha}}(\omega_{\alpha}\!-\!\omega_{\beta})
  \!\!\!&\equiv&\!\! \!-\!
  \int_{\theta_{\beta}} \!\!\frac{N(0)}{2\pi} \tilde{u}_0(\alpha,\beta) \label{eq:gamma}
\end{eqnarray}
 with $\eta^{\gamma}(\theta_{\beta},\theta) = \sum_{p=1}^{\infty}
\phi_p^{\gamma}(\theta_{\beta}) \phi_p^{\gamma}(\theta)$. 
The $u_0$ term that appears in the definition of $\bar{\Gamma}_{\theta_{\alpha}}$ via $\tilde{u}_0$, does not contribute to $Z$ because it does not have any dependence on frequency, and therefore the Matsubara sum in Eq. (\ref{eq:ge}) vanishes for this term.
The $u_0$ term in $ \Gamma^{\gamma}_{\theta,\theta^{\prime}}$ does however contribute to the Eq. (\ref{eq:ge}) involving $v^{\gamma}(\omega_n,\theta)$. There is one such generalized Eliashberg equation for each channel $\gamma$, which depend on the $\gamma$-component of the initial coupling through $\Gamma^{\gamma}_{\theta,\theta^{\prime}}$ as defined in Eq. (\ref{eq:gamma}). However, each of these equations, corresponding to a particular channel $\gamma$, also depends on the quasiparticle weight $Z$ which is renormalized by {\it all the components} of the initial coupling through $\bar{\Gamma}_{\theta}$ as defined in Eq. (\ref{eq:gamma}).
 
From the generalized Eliashberg equations, one can obtain the value of $T_c$
in terms of the microscopic parameters. For the most general case, analytical
expressions for $T_c$ are quite involved. We consider first the special case in which 
$g(\theta_{\alpha},\theta_{\beta})$ is separable, so that it has the
form $g(\theta_{\alpha},\theta_{\beta}) = f(\theta_{\alpha})
f(\theta_{\beta})$, and set $u_0=0$ for now. Then
$\Gamma^{\gamma}_{\theta,\theta^{\prime}}$ will also be
separable:
\begin{eqnarray}
\Gamma^{\gamma}_{\theta,\theta^{\prime}}
(\omega_{\alpha}-\omega_{\beta}) =
\Gamma_{\theta}^{\gamma}(\omega_{\alpha}-\omega_{\beta})
\Gamma_{\theta^{\prime}}^{\gamma}(\omega_{\alpha}-\omega_{\beta}), 
\end{eqnarray}
where
\begin{eqnarray}
\Gamma_{\theta}^{\gamma}(\omega_{\alpha}-\omega_{\beta})\equiv
\int_{\theta_{\alpha}} f^2(\theta_{\alpha}) \sqrt{2
N(0)D(\omega_{\alpha}-\omega_{\beta})}\eta^{\gamma}(\theta_{\alpha},\theta)\nonumber.
\end{eqnarray}
On the other hand, it is always possible to write:
\begin{eqnarray}
\bar{\Gamma}_{\theta}(\omega_{\alpha}-\omega_{\beta}) =
\frac{1}{2\pi}\int_{\theta^{\prime}}
\bar{\Gamma}_{\theta}(\omega_{\alpha}-\omega_{\beta})
\bar{\Gamma}_{\theta^{\prime}}(\omega_{\alpha}-\omega_{\beta}),
\end{eqnarray}
where
$\bar{\Gamma}_{\theta}(\omega_{\alpha}-\omega_{\beta}) \equiv
\sum_{\gamma}
\Gamma^{\gamma}_{\theta}(\omega_{\alpha}-\omega_{\beta})$. 
In  the general case, since we know that $g(\theta_{\alpha},\theta_{\beta}) =
g(\theta_{\beta},\theta_{\alpha})$, we can use the decomposition
$g(\theta_{\alpha},\theta_{\beta}) = \sum_i f_i(\theta_{\alpha})
f_i(\theta_{\beta})$, and therefore:
\begin{eqnarray}
\Gamma^{\gamma}_{\theta,\theta^{\prime}}
(\omega_{\alpha}-\omega_{\beta}) = \sum_i \Gamma_{i
\theta}^{\gamma}(\omega_{\alpha}-\omega_{\beta}) \Gamma_{i
\theta^{\prime}}^{\gamma}(\omega_{\alpha}-\omega_{\beta})\nonumber\\
\bar{\Gamma}_{\theta}(\omega_{\alpha}\!-\!\omega_{\beta}) =
\frac{1}{2\pi}\sum_i\int_{\theta^{\prime}}\bar{\Gamma}_{i
\theta}(\omega_{\alpha}\!-\!\omega_{\beta})
\bar{\Gamma}_{i\theta^{\prime}}(\omega_{\alpha}\!-\!\omega_{\beta})
\end{eqnarray}
The set of equations derived here
reduce to those given by earlier treatment of Eliashberg
theory for anisotropic electron-boson interaction by Daams and
Carbotte\cite{DC81}, if
we consider a single separable channel,  $\gamma=1$ and $n=1$.

If for a given channel
$\gamma$, the gap is dominated by one component $s$ such that the SC gap can
be written as
$\Delta_{\gamma}(\theta) = \Delta_0 \phi^{\gamma}_s(\theta)$, then the
corresponding McMillan
expression\cite{McM68} for $T_c$ becomes:
\begin{eqnarray}\label{MM-Eq}
T_c^{\gamma} \approx \omega_{E} \exp\left\{\frac{Z_{\gamma}}{\mu^*
\delta_{\gamma
1} - \lambda_{\gamma}}\right\}
\end{eqnarray}
where $\delta_{\gamma
1}$ restricts the contribution of $\mu^*$ to the channel $\gamma\!=\!1$ ($u_0$ 
only has first harmonic of $s$-wave component), $Z_{\gamma}^{-1} = \int_{\theta}
 \phi_q^{\gamma}(\theta)/[1+\bar{\lambda}(\theta)]$ and
 \begin{eqnarray}
 \lambda_{\gamma} = Z_{\gamma} \sum_{p}
 \int_{\theta_{\alpha},\theta_{\beta},\theta}
 \frac{\lambda(\theta_{\alpha},\theta_{\beta}) \phi_p^{\gamma}(\theta_{\alpha})
 \eta^{\gamma}(\theta_{\beta},\theta)
 \phi_q^{\gamma}(\theta)}{1+\bar{\lambda}(\theta)} \ ,\label{eq:eff}
\end{eqnarray}
with $\lambda(\theta_{\alpha},\theta_{\beta})\!\!\equiv\!\!2N(0)
 g^2(\theta_{\alpha},\theta_{\beta})/\omega_E$, and
 $\bar{\lambda}(\theta)\equiv (2\pi)^{-1}\int_{\theta^{\prime}}
 \lambda(\theta,\theta^{\prime})$. The corresponding Allen-Dynes
expression is 
\begin{equation}\label{AD-Eq}
T_c \approx \sqrt{\bar{\lambda}_{\gamma}} \omega_E,
\end{equation} where
\begin{equation}
\bar{\lambda}_{\gamma} \equiv \sum_{p}
\int_{\theta_{\alpha},\theta_{\beta},\theta}
\lambda(\theta_{\alpha},\theta_{\beta}) \phi_p^{\gamma}(\theta_{\alpha})
\eta^{\gamma}(\theta_{\beta},\theta)
\phi_q^{\gamma}(\theta).
\end{equation}
The effective electron-phonon parameter $\lambda_{\gamma}$ [Eq. (\ref{eq:eff})] that appears in the McMillan and Allen-Dynes equations for a particular channel $\gamma$ does include contributions from {\it all components} of the bare coupling $\lambda(\theta_{\alpha},\theta_{\beta})$ through $Z_{\gamma}$ and $\bar{\lambda}(\theta)$.
      
\section{Numerical results}

The method used allows
for an easy numerical calculation of the critical
temperature $T_c$ (as well as the zero temperature gap $\Delta_0$) from
the microscopic parameters $u_0$, $\lambda(\theta_{\alpha},\theta_{\beta})$ and $\omega_E$.
Eq. (\ref{U(l)}) gives the RG evolution of all the vertices and can be
evaluated at different RG steps $\ell$. To find the instability, one can
simply calculate the quantity $\det[\mathbf{1} + \mathbf{U}^{\gamma}(0)
  \cdot \mathbf{P}^{\gamma}(\ell)]$ and see when it approaches zero. Here we study electron-boson couplings with different angular dependences and strengths, going from the weak to the strong coupling regimes. We label $\theta$ the angle associated to ${\bf k}\equiv {\bf k}_1=-{\bf k}_2$ in polar coordinates, and $\theta'$ the angle of ${\bf k'}\equiv {\bf k}_3=-{\bf k}_4$. In general we define $g(\theta,\theta')=g_0f(\theta,\theta')$, where $g_0$ is a position independent constant and the function $f(\theta,\theta')$ 
contains the full angular dependence of the electron-boson matrix elements. We consider two different cases: a) The electron-boson coupling only depends on the difference between the angles of the electrons. b) The coupling depends on the exact position of each of the the electrons on the Fermi surface.

\subsection{Solution of the RG flow equations for $g(\theta,\theta')=g_0\cos(\theta-\theta')$}

\begin{figure}[t]
\resizebox{4.6cm}{!}
{\centerline{\includegraphics*{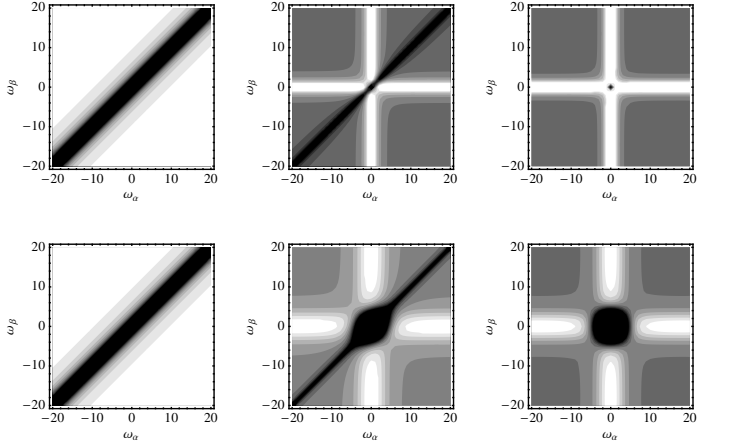}
}}
\caption{Plots of the $(2N+1)M \times (2N+1)M$ matrix
$\mathbf{U}^{\gamma}(\ell)$ at different RG scales $\ell$ for the $s$-channel
($\gamma=1$), for the electron-boson coupling $g(\theta,\theta')=g_0\cos(\theta-\theta')$. Here, for the
first harmonic in the expansion ($M=1$), the number of frequency
divisions is 41 ($N=20$) and
$\Lambda_0=100$, $\omega_E=10$, $u_0=0.1$. The three panels on
the top correspond to $\lambda=0.4$ (weak coupling) and the three on the
bottom correspond to $\lambda=4$ (strong coupling).}
\label{Evol-l-cos(th-th')1}
\end{figure}

For this kind of angular dependence, the strength of the electron-boson coupling does not depend on the specific position of the particles on the FS, but only on their relative orientation.
By solving Eq. (\ref{det-eq}) we obtain the evolution of
$\mathbf{U^{\gamma}}(\ell)$ with $\ell$.
For the coupling $g(\theta,\theta')=g_0\cos(\theta-\theta')$, only three channels contribute, corresponding to
$s$-, $d_{x^2-y^2}$- and $d_{xy}$-symmetry, respectively. Furthermore, the contributions of the two channels with $d$-symmetry have the same magnitudes, the channels are degenerate, and we will generically refer to them as $d$-wave channel. 
In Fig. \ref{Evol-l-cos(th-th')1} the evolution of $\mathbf{U^{\gamma}}(\ell)$ with $\ell$ for the
$s$-channel is depicted. Each panel represents the $(2N+1)M \times (2N+1)M$
metric $\mathbf{U}^{s}(\ell)$ at a given RG step $\ell$. The matrix
elements corresponding to small frequencies, around $\omega=0$, are at the center of
the panels. The three top panels of Fig. \ref{Evol-l-cos(th-th')1} show the RG flow evolution for a case in the weak coupling
regime ($\lambda = 0.4$).
The first panel on the left shows the initial condition, for which $\ell =0$. The third one, on the right hand side, shows the vertices at $\ell=\ell_c$,
right before the instability. We see that the most divergent couplings are those
with frequencies below the Einstein frequency,
$|\omega_{\alpha}|,|\omega_{\beta}|<\omega_E$. In this case, simple two-step
RG can be applied. The three panels on the bottom of
Fig. \ref{Evol-l-cos(th-th')1} represent the evolution of the
couplings for the strong electron-boson interaction regime ($\lambda=4$). In this
case, the most divergent couplings (at $\ell=\ell_c$) are those with frequencies
below a given energy scale $W_c > \omega_E$.
As expected, these results are similar to that obtained in Ref. \cite{Shan05b} where an isotropic electron-boson
coupling was considered. Furthermore, this can be seen as a proof of the validity of our method, which matches the results of the isotropic limit when only the $s$-channel is considered.

\begin{figure}[t]
\resizebox{4.6cm}{!}
{\centerline{\includegraphics*{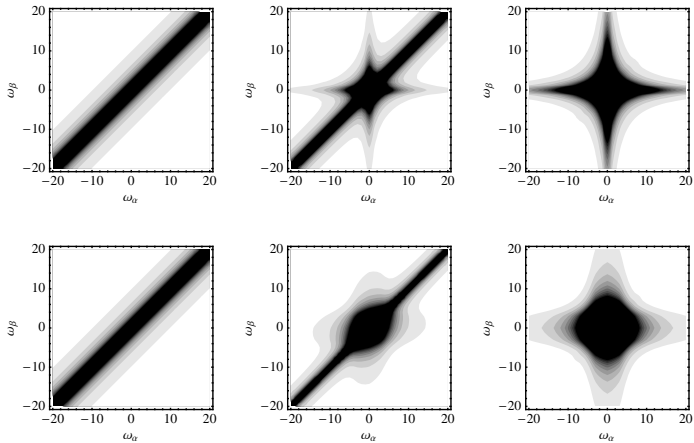}
}}
\caption{Same as Fig. \ref{Evol-l-cos(th-th')1} but for the
$d$-channel ($\gamma=2,3$). The three panels on the top are for weak
coupling ($\lambda=0.4$)
and the ones on the bottom are for the strong
coupling regime ($\lambda=4$).} \label{Evol-l-cos(th-th')2}
\end{figure}

\begin{figure}[b]
\resizebox{4.8cm}{!}
{\centerline{\includegraphics*{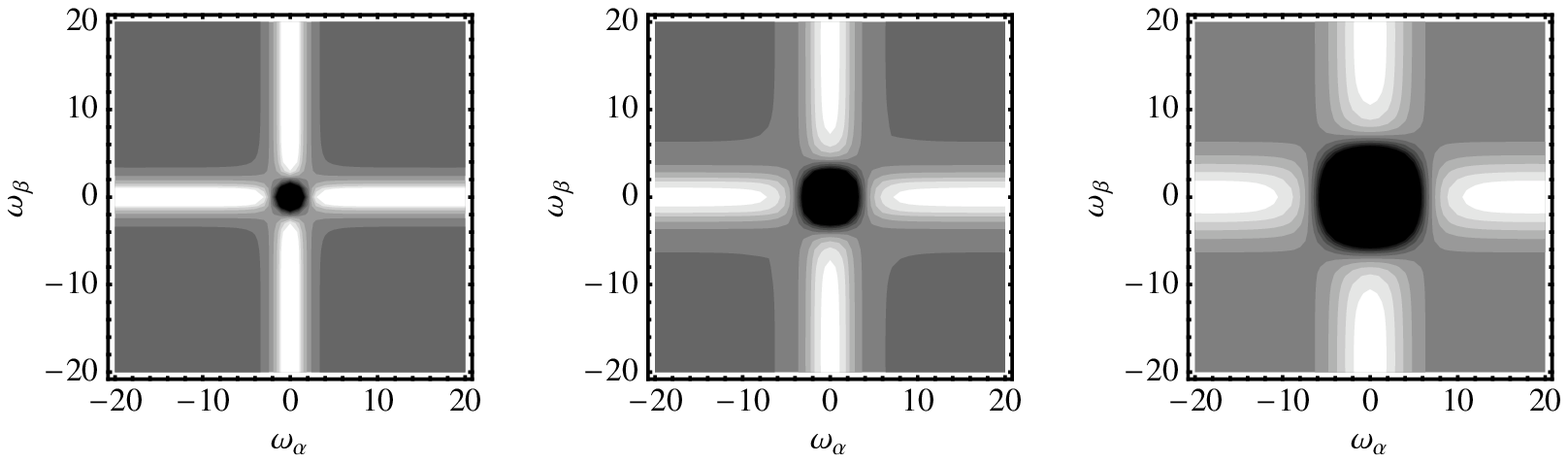}
}}
\caption{Coupling matrix $\mathbf{U}^s(\ell_c)$ at
the instability point for the $s$-channel, from weak
to strong coupling, for $u_0=0.1$. The corresponding electron-boson coupling is $g(\theta,\theta')=g_0\cos(\theta-\theta')$. Panels correspond, from left to right, to $\lambda=
1, 2.5, 6$.} \label{s-evolution}
\end{figure}

\begin{figure}[b]
\resizebox{4.8cm}{!}
{\centerline{\includegraphics*{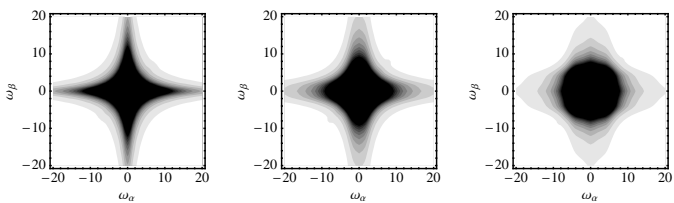}
}}
\caption{Same as Fig. \ref{s-evolution} and for the same parameter values, but for the $d$-channel, $\mathbf{U}^d(\ell_c)$.} \label{d-evolution}
\end{figure}

The situation is quite different for the $d$-channel, as it can be seen in Fig. \ref{Evol-l-cos(th-th')2}. Again, the
panels on the top corresponds to the weak coupling regime. The initial condition is the same as for the $s$-channel, with the $u^d(\omega,-\omega,\ell=0)$ being the largest couplings. However, close to the instability, the couplings
$u^d(\omega_{\alpha},\omega_{\beta},\ell_c)$ that first diverge
correspond to a region of frequencies centered around $\omega=0$ and with a
star-like shape. This frequency dependence of $u^d$ at the instability point illustrates the importance of retardation effects in the RG flow. To have a physical understanding of this specific structure is, however, not trivial, and we do not speculate here about the possible physical consequences of such dependence. On the other hand, the bottom panels of Fig. \ref{Evol-l-cos(th-th')2}
represents the flow, in the strong coupling regime, from the
initial condition $\ell=0$, to the critical RG step $\ell_c$. In this case, the most
divergent couplings at the instability point are, as in the $s$-wave case, those with
frequencies below $W_c$.

In Fig. \ref{s-evolution} we show the density plots of the
matrix $\mathbf{U}^s(\ell_c)$ at the
RG  scale $\ell_c$, where the instability appears, for different values of the electron-boson coupling strength.
At this point  we can see more
clearly how the energy scale separating high and low energy
physics moves from the Einstein frequency $\omega_E$, in the weak
coupling regime, to the critical cutoff $W_c$, in the strong
coupling regime. The scale $W_c$ can be associated with the $T=0$ superconducting
gap $\Delta_0$, or with the critical temperature 
$T_c$ of the SC phase in the finite
temperature formalism\cite{Shan05,Shan05b}. 
Fig. \ref{s-evolution} also illustrates 
the breakdown of the two-step RG. In this approximation, the
vertex is chosen to be just the electron-electron part for
frequencies above the Einstein frequency, and to have some
constant contribution from the boson modes for frequencies below
$\omega_E$ \cite{P94}. This approximation works well in the weak
coupling limit, where the most divergent couplings are those with
frequencies below $\omega_E$. But for large $\lambda$ ($\lambda \gg 1$) this behavior
breaks down and the scale
for the divergent central region is of order $W_c > \omega_E$.

\begin{figure}[t]
\resizebox{8.5cm}{!}{\centerline{\includegraphics[]{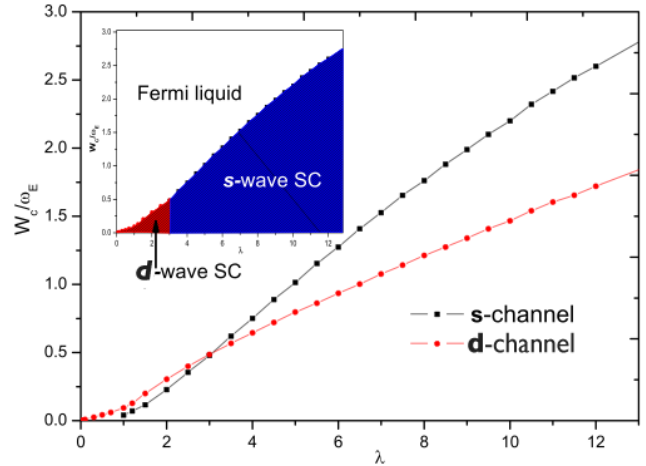}}}
\caption{(Color online) Critical cutoff $W_c$ versus $\lambda$ for the
$s$-channel (black squares) and the $d$-channel (red
circles). The value of the parameters used are $\Lambda_0=100$,
$\omega_E=10$ and $u_0=2.0$. The inset represents the phase
diagram.} \label{Wc-lambda}
\end{figure}

Similarly,  the couplings $\mathbf{U}^d(\ell_c)$ 
in the $d$-channel are shown in Fig. \ref{d-evolution}, from the weak to the strong
electron-boson  coupling regimes. We
obtain the star-like structure at weak coupling ($\lambda=1$),
evolving towards a more circular form in the strong
coupling ($\lambda=6$). More analytical study is required for a full understanding of these patterns.

Fig. \ref{Wc-lambda} shows the energy scale $W_c$, where the SC instability
occurs, as a function of the
strength of electron-boson coupling $\lambda$, at fixed $\omega_E$. The two channels  ($s$ and $d$) that contribute to the electron-boson
coupling $g(\theta,\theta')=g_0\cos(\theta-\theta')$, are
represented. The behavior of $W_c$ in the weak coupling limit follows the McMillan exponential behavior of Eq. (\ref{MM-Eq}).
For large $\lambda$, strong coupling regime, the  $W_c$ follows the  Allen-Dynes law, Eq. (\ref{AD-Eq}). In the inset of Fig. \ref{Wc-lambda} we show  the
phase diagram of the system. At finite temperature,
the Fermi liquid  phase  breaks down towards
$d$-wave SC in the weak to intermediate coupling
range, and towards $s$-wave SC in the intermediate to
strong coupling regime. The larger the on-site
electronic repulsion $u_0$, the larger the electron-boson 
coupling crossover between $s$- and $d$-wave superconductivity. This is expected since an isotropic
$u_0$ only suppresses the contribution to the $s$-wave channel, without affecting the other
channels. Therefore for $u_0=0$, superconductivity only occurs in the $s$-channel. Notice that, because the Kohn-Luttinger effect is not included in this one-loop RG calculation, there is no superconducting instability when only electron-electron repulsion is considered. In addition, this effect should shift the transition between $d$-wave and $s$-wave superconductivity toward some different value of the electron-boson coupling, $\lambda$. The qualitative behavior is, however, well captured by our approximation, as shown in the phase diagram of Fig. \ref{Wc-lambda}.

\subsection{Solution of the RG flow equations for $g(\theta,\theta')=g_0 \cos(\theta/2) \cos(\theta'/2)$}

\begin{figure}[t]
\resizebox{4.5cm}{!}{\centerline{\includegraphics[]{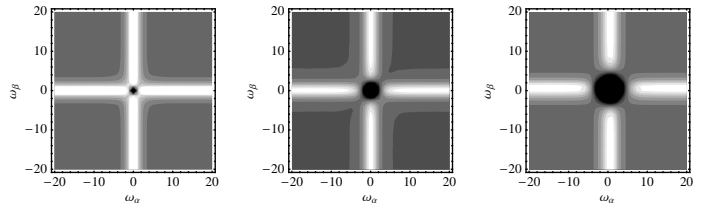}}}
\caption{Coupling matrix $\mathbf{U}^s(\ell_c)$ at
the instability point for the $s$-channel, from weak
to strong coupling, for $u_0=0.1$. The corresponding electron-boson coupling is $g(\theta,\theta')=g_0\cos(\theta/2)\cos(\theta'/2)$. Panels correspond, from left to right, to $\lambda=
1, 2.5, 6$.} \label{s-u0-0.1}
\end{figure}

In the following we  consider an electron-boson coupling with the angular dependence
$g(\theta,\theta')=g_0 \cos(\theta/2) \cos(\theta'/2)$. Unlike the previous case, this is a simple example for which the boson mode couples differently to electrons in different parts of the Fermi surface. That is, the coupling depends on the specific position of each of the electrons in the Fermi surface. As before, we obtain the $\ell$-evolution of $\mathbf{U^{\gamma}}(\ell)$ by solving Eq. (\ref{det-eq}) for this coupling. The two contributing channels in this case are
of $s$- and $p$-symmetry. The evolution of $\mathbf{U}^{p}(\ell)$ for the $p$-channel, at weak electron-boson
coupling, shows a star-shaped structure, as it was found in the preceding section 
for the $d$-channel. The frequency dependence of the coupling matrix close
to the instability is very similar to that represented in the top right panel of Fig. \ref{Evol-l-cos(th-th')2}.

\begin{figure}[b]
\resizebox{4.5cm}{!}{\centerline{\includegraphics[]{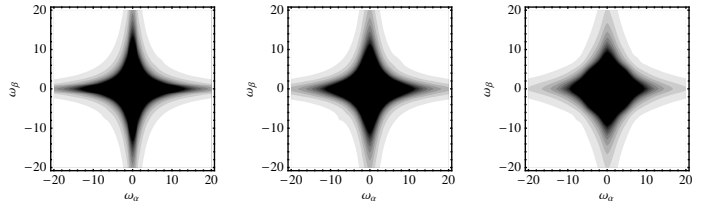}}}
\caption{Same as Fig. \ref{s-u0-0.1} and for the same parameter values, but for the $p$-channel, $\mathbf{U}^p(\ell_c)$.} \label{p-evol}
\end{figure}

  In Fig. \ref{s-u0-0.1}, the density plots of $\mathbf{U}^{s}(\ell_c)$
are shown for $u_0=0.1$ and different values of $\lambda$, at the critical RG step $\ell_c$. They present the same qualitative behavior as the corresponding $s$-channel results of the previous subsection (see Fig. \ref{d-evolution}).
For the $p$-channel, the density plots of the coupling matrix $\mathbf{U}^{p}(\ell_c)$ at
the instability point $\ell_c$, are depicted in Fig. \ref{p-evol}. The plot for $\lambda= 1.0$ is very similar to that obtained for the   $d$-channel,  in the case of a coupling of the form $g\propto \cos(\theta-\theta')$. This can be observed by comparing the left hand side graphs of Fig. \ref {d-evolution} and \ref{p-evol}. However, the situation is different in the strong coupling limit. The corresponding plot for the $d$-channel shows that
the most diverging matrix elements lie in a circular shaped region around $\omega=0$ (see the right panel
of Fig. \ref {d-evolution}). However in the present case, where $g\propto  \cos(\theta/2) \cos(\theta'/2)$, the diverging matrix
elements for the $p$-channel are confined in a square region around $\omega=0$, as shown in Fig. \ref {p-evol}. This different behavior
in the strong coupling region needs a better understanding.

Finally, we mention that it is possible to build a phase diagram similar to that of Fig. \ref{Wc-lambda}, but for a coupling of the form $g\propto  \cos(\theta/2) \cos(\theta'/2)$. In this case, instead of having a low-coupling $d$-wave SC region in the phase diagram, we obtain a zone with $p$-wave pairing SC. For a value of $u_0=0.5$, the crossover between $p$- and $s$-wave superconductivity occurs at $\lambda\sim 2.5$. For a small value of the on-site Coulomb interaction, $u_0=0.1$, it is found that the Fermi liquid state is unstable towards $p$-wave SC in the range $0<\lambda\lesssim 0.7$, and towards $s$-wave SC for larger values of the coupling, $\lambda \gtrsim 0.7$.

\section{Conclusions}

In this work, we have  investigated the pairing instabilities of the FL state which appear 
when the electron-boson coupling overcomes the effective repulsive electron-electron interaction. The most general case of anisotropic coupling of electrons
to bosons has been considered. The
anisotropic boson exchange couplings are treated by  an RG approach.
Calculations are made for a 2D square lattice at low fillings. Under this assumption, the system presents an almost circular FS, which allows for an analytical solution of the RG flow equations. Furthermore, the numerical evaluation of the flows provides information of the change
of the frequency dependence of the vertices. The flow equations 
for the BCS vertex are decomposed into contributions coming from different symmetry channels, with different angular
momentum dependence. By varying the strength of the interaction, from the weak to the strong coupling
regime, the evolution of the couplings for the different symmetry channels is obtained.
Channels of $s$- , $p$-, and $d$-symmetry have been investigated.

 We have considered here  simple functional forms of the electron-boson coupling, such as 
$g(\theta,\theta')=g_0\cos(\theta-\theta')$, where the momentum of the exchanged boson only depends
on the angle difference between the two electrons involved, and 
$g(\theta,\theta')=g_0 \cos(\theta/2) \cos(\theta'/2)$, where the coupling depends explicitly on the position of each of the electrons on the Fermi surface. It is found that, even  with these simple angular dependences,
the anisotropic electron-boson couplings induce new non-trivial physics. As far as we know, the inclusion of angular dependence of the electron-phonon coupling
has not been addressed before in previous RG calculations. Although academic at first view, this problem makes contact with the physics 
which appears in some new anisotropic materials as those described in the Introduction.
In these materials, metal-transition borocarbides, boronitrides, magnesium diboride,
or cuprates, standard BCS theory cannot explain many of their properties. A consistent description of some physical behaviors is however achieved
if anisotropy of the electron-boson coupling is included in the Eliashberg 
theory \cite{MMEHS01,C90,DCSN04}. In high-temperature superconductors, there is a considerable 
amount of data that point to interplay between electronic and atomic
degrees of freedom \cite{CYGZ08}. The different behavior shown by the quasiparticles
in both nodal and antinodal regions of the Brillouin zone, has suggested, among other
explanations, an anisotropic electron-phonon coupling, along side other many-body effects, in order to understand the pairing mechanism \cite{MKD08}. Coupling to the half-breathing mode in the nodal region and to the buckling mode in the antinodal direction have
been propose as an interpretation to the renormalization effects seen in ARPES results in
both, the normal and superconducting phases \cite{DCSN04}. However, more work is needed to get insight in the complex 
interplay between electron-boson interactions and electronic correlation in unconventional superconductivity.

In summary, we have studied the superconducting instability of the Fermi liquid phase, considering electron-boson couplings with different angular dependences and strengths. SC order parameters of $s$, $p$ and $d$ symmetry have been obtained, depending on the anisotropy of the coupling and the strength of the interaction.
The  investigation of the frequency dependence of the 
couplings of different symmetries could be interesting in order to
analyze and understand the complex behavior revealed by the experiments.
Therefore, numerical evaluation of the RG flow equations
are instructive in determining the range of important frequencies in
different regimes and for different symmetry channels. Furthermore, at the instability point and for finite temperatures, the RG equations give the 
solution of the generalized Eliashberg equations at $T_c$ and consequently, McMillan and Allen-Dynes expressions have been obtained.

\acknowledgments
Funding from MCyT (Spain) through grants FIS2005-05478-C02-01and FIS2008-00124 is acknowledged. 
R.R. appreciates the hospitality of the UCR, where part of this work was done. R.R. acknowledges financial funding from the Agence Nationale de la Recherche under Grant No. ANR-06-NANO-019-03 and from the RTRA "Triangle de la Physique".
We appreciate useful conversations with F. D. Klironomos and A. H. Castro Neto.

\bibliography{BibliogrAnisotropicRG}

\newcommand{\npb}{Nucl. Phys.}\newcommand{\adv}{Adv.
  Phys.}\newcommand{\epl}{Europhys. Lett.}
\begin{thebibliography}{46}
\expandafter\ifx\csname natexlab\endcsname\relax\def\natexlab#1{#1}\fi
\expandafter\ifx\csname bibnamefont\endcsname\relax
  \def\bibnamefont#1{#1}\fi
\expandafter\ifx\csname bibfnamefont\endcsname\relax
  \def\bibfnamefont#1{#1}\fi
\expandafter\ifx\csname citenamefont\endcsname\relax
  \def\citenamefont#1{#1}\fi
\expandafter\ifx\csname url\endcsname\relax
  \def\url#1{\texttt{#1}}\fi
\expandafter\ifx\csname urlprefix\endcsname\relax\def\urlprefix{URL }\fi
\providecommand{\bibinfo}[2]{#2}
\providecommand{\eprint}[2][]{\url{#2}}

\bibitem[{\citenamefont{Eliashberg}(1960)}]{E60}
\bibinfo{author}{\bibfnamefont{G.~M.} \bibnamefont{Eliashberg}},
  \bibinfo{journal}{Zh. Exp. Teor. Fiz.} \textbf{\bibinfo{volume}{38}},
  \bibinfo{pages}{966} (\bibinfo{year}{1960}).

\bibitem[{\citenamefont{Carbotte}(1990)}]{C90}
\bibinfo{author}{\bibfnamefont{J.~P.} \bibnamefont{Carbotte}},
  \bibinfo{journal}{Rev. Mod. Phys.} \textbf{\bibinfo{volume}{62}},
  \bibinfo{pages}{1027} (\bibinfo{year}{1990}).

\bibitem[{\citenamefont{Morel and Anderson}(1962)}]{MA62}
\bibinfo{author}{\bibfnamefont{P.}~\bibnamefont{Morel}} \bibnamefont{and}
  \bibinfo{author}{\bibfnamefont{P.~W.} \bibnamefont{Anderson}},
  \bibinfo{journal}{Phys. Rev.} \textbf{\bibinfo{volume}{125}},
  \bibinfo{pages}{1263} (\bibinfo{year}{1962}).

\bibitem[{\citenamefont{McMillan and Rowell}(1965)}]{McMR65}
\bibinfo{author}{\bibfnamefont{W.~L.} \bibnamefont{McMillan}} \bibnamefont{and}
  \bibinfo{author}{\bibfnamefont{J.~M.} \bibnamefont{Rowell}},
  \bibinfo{journal}{Phys. Rev. Lett.} \textbf{\bibinfo{volume}{14}},
  \bibinfo{pages}{108} (\bibinfo{year}{1965}).

\bibitem[{\citenamefont{McMillan and { Rowell, in {\it Superconductivity}, Vol.
  1, edited by R. D. Parks}}(Dekker, New York, 1969)}]{McMR69}
\bibinfo{author}{\bibfnamefont{W.~L.} \bibnamefont{McMillan}} \bibnamefont{and}
  \bibinfo{author}{\bibfnamefont{J.~M.} \bibnamefont{{ Rowell, in {\it
  Superconductivity}, Vol. 1, edited by R. D. Parks}}} (\bibinfo{year}{Dekker,
  New York, 1969}).

\bibitem[{\citenamefont{{Scalapino, in {\it Superconductivity}, Vol. 1, edited
  by R. D. Parks}}(Dekker, New York, 1969)}]{Scal69}
\bibinfo{author}{\bibfnamefont{D.~J.} \bibnamefont{{Scalapino, in {\it
  Superconductivity}, Vol. 1, edited by R. D. Parks}}} (\bibinfo{year}{Dekker,
  New York, 1969}).

\bibitem[{\citenamefont{McMillan}(1968)}]{McM68}
\bibinfo{author}{\bibfnamefont{W.~L.} \bibnamefont{McMillan}},
  \bibinfo{journal}{Phys. Rev.} \textbf{\bibinfo{volume}{167}},
  \bibinfo{pages}{331} (\bibinfo{year}{1968}).

\bibitem[{\citenamefont{Manalo et~al.}(2001)\citenamefont{Manalo, Michor,
  Hagary, Hilscher, and Schachinger}}]{MMEHS01}
\bibinfo{author}{\bibfnamefont{S.}~\bibnamefont{Manalo}},
  \bibinfo{author}{\bibfnamefont{M.}~\bibnamefont{Michor}},
  \bibinfo{author}{\bibfnamefont{E.~E.} \bibnamefont{Hagary}},
  \bibinfo{author}{\bibfnamefont{G.}~\bibnamefont{Hilscher}}, \bibnamefont{and}
  \bibinfo{author}{\bibfnamefont{E.}~\bibnamefont{Schachinger}},
  \bibinfo{journal}{Phys. Rev. B} \textbf{\bibinfo{volume}{63}},
  \bibinfo{pages}{104508} (\bibinfo{year}{2001}).

\bibitem[{\citenamefont{Manalo and Schachinger}(2001)}]{MS01}
\bibinfo{author}{\bibfnamefont{S.}~\bibnamefont{Manalo}} \bibnamefont{and}
  \bibinfo{author}{\bibfnamefont{E.}~\bibnamefont{Schachinger}},
  \bibinfo{journal}{J. Low Temp. Phys.} \textbf{\bibinfo{volume}{123}},
  \bibinfo{pages}{149} (\bibinfo{year}{2001}).

\bibitem[{\citenamefont{Buzea and Yamashita}(2001)}]{BY01}
\bibinfo{author}{\bibfnamefont{C.}~\bibnamefont{Buzea}} \bibnamefont{and}
  \bibinfo{author}{\bibfnamefont{T.}~\bibnamefont{Yamashita}},
  \bibinfo{journal}{Supercond. Sci. Technol.} \textbf{\bibinfo{volume}{14}},
  \bibinfo{pages}{R115} (\bibinfo{year}{2001}).

\bibitem[{\citenamefont{Nagamatsu et~al.}(2001)\citenamefont{Nagamatsu,
  Nakagawa, Muranaka, Zenitani, and Akimitsu}}]{NNMZA01}
\bibinfo{author}{\bibfnamefont{J.}~\bibnamefont{Nagamatsu}},
  \bibinfo{author}{\bibfnamefont{N.}~\bibnamefont{Nakagawa}},
  \bibinfo{author}{\bibfnamefont{T.}~\bibnamefont{Muranaka}},
  \bibinfo{author}{\bibfnamefont{Y.}~\bibnamefont{Zenitani}}, \bibnamefont{and}
  \bibinfo{author}{\bibfnamefont{J.}~\bibnamefont{Akimitsu}},
  \bibinfo{journal}{Nature} \textbf{\bibinfo{volume}{410}}, \bibinfo{pages}{63}
  (\bibinfo{year}{2001}).

\bibitem[{\citenamefont{Allen and B.{ Mitrovic, in {\it Solid State Physics},
  Vol.37, edited by H. Ehrenreich and F. Seit and D. Turnbull}}(Academic, New
  York, 1982)}]{AM82}
\bibinfo{author}{\bibfnamefont{P.~B.} \bibnamefont{Allen}} \bibnamefont{and}
  \bibinfo{author}{\bibnamefont{B.{ Mitrovic, in {\it Solid State Physics},
  Vol.37, edited by H. Ehrenreich and F. Seit and D. Turnbull}}}
  (\bibinfo{year}{Academic, New York, 1982}).

\bibitem[{\citenamefont{Choi et~al.}(2002)\citenamefont{Choi, Roundy, Sun,
  Cohen, and Louie}}]{CRSCL02}
\bibinfo{author}{\bibfnamefont{H.~J.} \bibnamefont{Choi}},
  \bibinfo{author}{\bibfnamefont{D.}~\bibnamefont{Roundy}},
  \bibinfo{author}{\bibfnamefont{H.}~\bibnamefont{Sun}},
  \bibinfo{author}{\bibfnamefont{M.}~\bibnamefont{Cohen}}, \bibnamefont{and}
  \bibinfo{author}{\bibfnamefont{S.}~\bibnamefont{Louie}},
  \bibinfo{journal}{Nature} \textbf{\bibinfo{volume}{418}},
  \bibinfo{pages}{758} (\bibinfo{year}{2002}).

\bibitem[{\citenamefont{Millis et~al.}(1988)\citenamefont{Millis, Sachdev, and
  Varma}}]{MSV88}
\bibinfo{author}{\bibfnamefont{A.~J.} \bibnamefont{Millis}},
  \bibinfo{author}{\bibfnamefont{S.}~\bibnamefont{Sachdev}}, \bibnamefont{and}
  \bibinfo{author}{\bibfnamefont{C.~M.} \bibnamefont{Varma}},
  \bibinfo{journal}{Phys. Rev. B} \textbf{\bibinfo{volume}{37}},
  \bibinfo{pages}{4975} (\bibinfo{year}{1988}).

\bibitem[{\citenamefont{Balatsky and Zhu}(2006)}]{BZ06}
\bibinfo{author}{\bibfnamefont{A.~V.} \bibnamefont{Balatsky}} \bibnamefont{and}
  \bibinfo{author}{\bibfnamefont{J.-X.} \bibnamefont{Zhu}},
  \bibinfo{journal}{Phys. Rev. B} \textbf{\bibinfo{volume}{74}},
  \bibinfo{pages}{094517} (\bibinfo{year}{2006}).

\bibitem[{\citenamefont{Millis}(1992)}]{M92}
\bibinfo{author}{\bibfnamefont{A.~J.} \bibnamefont{Millis}},
  \bibinfo{journal}{Phys.Rev.B} \textbf{\bibinfo{volume}{45}},
  \bibinfo{pages}{13047} (\bibinfo{year}{1992}).

\bibitem[{\citenamefont{Lanzara et~al.}(2001)\citenamefont{Lanzara, Bogdanov,
  Zhou, Kellar, Feng, Lu, Yoshida, Eisaki, Fujimori, Kishio et~al.}}]{L01}
\bibinfo{author}{\bibfnamefont{A.}~\bibnamefont{Lanzara}},
  \bibinfo{author}{\bibfnamefont{P.~V.} \bibnamefont{Bogdanov}},
  \bibinfo{author}{\bibfnamefont{X.~J.} \bibnamefont{Zhou}},
  \bibinfo{author}{\bibfnamefont{S.~A.} \bibnamefont{Kellar}},
  \bibinfo{author}{\bibfnamefont{D.~L.} \bibnamefont{Feng}},
  \bibinfo{author}{\bibfnamefont{E.~D.} \bibnamefont{Lu}},
  \bibinfo{author}{\bibfnamefont{T.}~\bibnamefont{Yoshida}},
  \bibinfo{author}{\bibfnamefont{H.}~\bibnamefont{Eisaki}},
  \bibinfo{author}{\bibfnamefont{A.}~\bibnamefont{Fujimori}},
  \bibinfo{author}{\bibfnamefont{K.}~\bibnamefont{Kishio}},
  \bibnamefont{et~al.}, \bibinfo{journal}{Nature}
  \textbf{\bibinfo{volume}{412}}, \bibinfo{pages}{510} (\bibinfo{year}{2001}).

\bibitem[{\citenamefont{Valla et~al.}(2007)\citenamefont{Valla, Kidd, Pan,
  Fedorov, Yin, Gu, and Johnson}}]{V06}
\bibinfo{author}{\bibfnamefont{T.}~\bibnamefont{Valla}},
  \bibinfo{author}{\bibfnamefont{T.}~\bibnamefont{Kidd}},
  \bibinfo{author}{\bibfnamefont{Z.-H.} \bibnamefont{Pan}},
  \bibinfo{author}{\bibfnamefont{A.}~\bibnamefont{Fedorov}},
  \bibinfo{author}{\bibfnamefont{W.-G.} \bibnamefont{Yin}},
  \bibinfo{author}{\bibfnamefont{G.}~\bibnamefont{Gu}}, \bibnamefont{and}
  \bibinfo{author}{\bibfnamefont{P.}~\bibnamefont{Johnson}},
  \bibinfo{journal}{Phys. Rev. Lett.} \textbf{\bibinfo{volume}{98}},
  \bibinfo{pages}{167003} (\bibinfo{year}{2007}).

\bibitem[{\citenamefont{Borisenko et~al.}(2003)\citenamefont{Borisenko,
  Kordyuk, Kim, Koitzsch, Knupfer, Fink, Golden, Eschrig, Berger, and
  Follath}}]{B03}
\bibinfo{author}{\bibfnamefont{S.~V.} \bibnamefont{Borisenko}},
  \bibinfo{author}{\bibfnamefont{A.~A.} \bibnamefont{Kordyuk}},
  \bibinfo{author}{\bibfnamefont{T.~K.} \bibnamefont{Kim}},
  \bibinfo{author}{\bibfnamefont{A.}~\bibnamefont{Koitzsch}},
  \bibinfo{author}{\bibfnamefont{M.}~\bibnamefont{Knupfer}},
  \bibinfo{author}{\bibfnamefont{J.}~\bibnamefont{Fink}},
  \bibinfo{author}{\bibfnamefont{M.~S.} \bibnamefont{Golden}},
  \bibinfo{author}{\bibfnamefont{M.}~\bibnamefont{Eschrig}},
  \bibinfo{author}{\bibfnamefont{H.}~\bibnamefont{Berger}}, \bibnamefont{and}
  \bibinfo{author}{\bibfnamefont{R.}~\bibnamefont{Follath}},
  \bibinfo{journal}{Phys. Rev. Lett.} \textbf{\bibinfo{volume}{90}},
  \bibinfo{pages}{207001} (\bibinfo{year}{2003}).

\bibitem[{\citenamefont{Honerkamp et~al.}(2007)\citenamefont{Honerkamp, Fu, and
  Lee}}]{HFL07}
\bibinfo{author}{\bibfnamefont{C.}~\bibnamefont{Honerkamp}},
  \bibinfo{author}{\bibfnamefont{H.}~\bibnamefont{Fu}}, \bibnamefont{and}
  \bibinfo{author}{\bibfnamefont{D.}~\bibnamefont{Lee}},
  \bibinfo{journal}{Phys. Rev. B} \textbf{\bibinfo{volume}{75}},
  \bibinfo{pages}{014503} (\bibinfo{year}{2007}).

\bibitem[{\citenamefont{Gweon et~al.}(2004)\citenamefont{Gweon, Sasagawa, Zhou,
  Graf, takagi, Lee, and Lanzara}}]{GSZGTLL04}
\bibinfo{author}{\bibfnamefont{G.-H.} \bibnamefont{Gweon}},
  \bibinfo{author}{\bibfnamefont{T.}~\bibnamefont{Sasagawa}},
  \bibinfo{author}{\bibfnamefont{S.}~\bibnamefont{Zhou}},
  \bibinfo{author}{\bibfnamefont{J.}~\bibnamefont{Graf}},
  \bibinfo{author}{\bibfnamefont{H.}~\bibnamefont{takagi}},
  \bibinfo{author}{\bibfnamefont{D.-H.} \bibnamefont{Lee}}, \bibnamefont{and}
  \bibinfo{author}{\bibfnamefont{A.}~\bibnamefont{Lanzara}},
  \bibinfo{journal}{Nature} \textbf{\bibinfo{volume}{430}},
  \bibinfo{pages}{187} (\bibinfo{year}{2004}).

\bibitem[{\citenamefont{Lee et~al.}(2006)\citenamefont{Lee, Fujita, McElroy,
  Slezak, Wang, Aiura, Bando, Ishikado, Masui, Zhu et~al.}}]{LFMS06}
\bibinfo{author}{\bibfnamefont{J.}~\bibnamefont{Lee}},
  \bibinfo{author}{\bibfnamefont{K.}~\bibnamefont{Fujita}},
  \bibinfo{author}{\bibfnamefont{K.}~\bibnamefont{McElroy}},
  \bibinfo{author}{\bibfnamefont{J.}~\bibnamefont{Slezak}},
  \bibinfo{author}{\bibfnamefont{M.}~\bibnamefont{Wang}},
  \bibinfo{author}{\bibfnamefont{Y.}~\bibnamefont{Aiura}},
  \bibinfo{author}{\bibfnamefont{H.}~\bibnamefont{Bando}},
  \bibinfo{author}{\bibfnamefont{M.}~\bibnamefont{Ishikado}},
  \bibinfo{author}{\bibfnamefont{T.}~\bibnamefont{Masui}},
  \bibinfo{author}{\bibfnamefont{J.-X.} \bibnamefont{Zhu}},
  \bibnamefont{et~al.}, \bibinfo{journal}{Nature}
  \textbf{\bibinfo{volume}{442}}, \bibinfo{pages}{546} (\bibinfo{year}{2006}).

\bibitem[{\citenamefont{Khsanov et~al.}(2008)\citenamefont{Khsanov, Strässle,
  Conder, Pomjakushina, Bussmann-Holder, and Keller}}]{KSCP07}
\bibinfo{author}{\bibfnamefont{R.}~\bibnamefont{Khsanov}},
  \bibinfo{author}{\bibfnamefont{S.}~\bibnamefont{Strässle}},
  \bibinfo{author}{\bibfnamefont{K.}~\bibnamefont{Conder}},
  \bibinfo{author}{\bibfnamefont{E.}~\bibnamefont{Pomjakushina}},
  \bibinfo{author}{\bibfnamefont{A.}~\bibnamefont{Bussmann-Holder}},
  \bibnamefont{and} \bibinfo{author}{\bibfnamefont{H.}~\bibnamefont{Keller}},
  \bibinfo{journal}{Phys. Rev. B} \textbf{\bibinfo{volume}{77}},
  \bibinfo{pages}{104530} (\bibinfo{year}{2008}).

\bibitem[{\citenamefont{Iwasawa et~al.}(2008)\citenamefont{Iwasawa, Douglas,
  Sato, Masui, Yoshida, Sun, Eisaki, Bando, Ino, Arita et~al.}}]{IDSM08}
\bibinfo{author}{\bibfnamefont{H.}~\bibnamefont{Iwasawa}},
  \bibinfo{author}{\bibfnamefont{J.~F.} \bibnamefont{Douglas}},
  \bibinfo{author}{\bibfnamefont{K.}~\bibnamefont{Sato}},
  \bibinfo{author}{\bibfnamefont{T.}~\bibnamefont{Masui}},
  \bibinfo{author}{\bibfnamefont{Y.}~\bibnamefont{Yoshida}},
  \bibinfo{author}{\bibfnamefont{Z.}~\bibnamefont{Sun}},
  \bibinfo{author}{\bibfnamefont{H.}~\bibnamefont{Eisaki}},
  \bibinfo{author}{\bibfnamefont{H.}~\bibnamefont{Bando}},
  \bibinfo{author}{\bibfnamefont{A.}~\bibnamefont{Ino}},
  \bibinfo{author}{\bibfnamefont{M.}~\bibnamefont{Arita}},
  \bibnamefont{et~al.}, \bibinfo{journal}{\prl} \textbf{\bibinfo{volume}{101}},
  \bibinfo{pages}{157005} (\bibinfo{year}{2008}).

\bibitem[{\citenamefont{Devereaux et~al.}(2004)\citenamefont{Devereaux, Cuk,
  Shen, and Nagaosa}}]{DCSN04}
\bibinfo{author}{\bibfnamefont{T.~P.} \bibnamefont{Devereaux}},
  \bibinfo{author}{\bibfnamefont{T.}~\bibnamefont{Cuk}},
  \bibinfo{author}{\bibfnamefont{Z.-X.} \bibnamefont{Shen}}, \bibnamefont{and}
  \bibinfo{author}{\bibfnamefont{N.}~\bibnamefont{Nagaosa}},
  \bibinfo{journal}{Phys. Rev. Lett.} \textbf{\bibinfo{volume}{93}},
  \bibinfo{pages}{117004} (\bibinfo{year}{2004}).

\bibitem[{\citenamefont{Tsai et~al.}(2005)\citenamefont{Tsai, {Castro Neto},
  Shankar, and Campbell}}]{Shan05}
\bibinfo{author}{\bibfnamefont{S.-W.} \bibnamefont{Tsai}},
  \bibinfo{author}{\bibfnamefont{A.~H.} \bibnamefont{{Castro Neto}}},
  \bibinfo{author}{\bibfnamefont{R.}~\bibnamefont{Shankar}}, \bibnamefont{and}
  \bibinfo{author}{\bibfnamefont{D.~K.} \bibnamefont{Campbell}},
  \bibinfo{journal}{Phys. Rev. B} \textbf{\bibinfo{volume}{72}},
  \bibinfo{pages}{054531} (\bibinfo{year}{2005}).

\bibitem[{\citenamefont{Migdal}(1958)}]{Migdal58}
\bibinfo{author}{\bibfnamefont{A.~B.} \bibnamefont{Migdal}},
  \bibinfo{journal}{Zh. Eksp. Teor. Fiz.} \textbf{\bibinfo{volume}{34}},
  \bibinfo{pages}{1438} (\bibinfo{year}{1958}).

\bibitem[{\citenamefont{Tam et~al.}(2007)\citenamefont{Tam, S.-W.Tsai,
  Campbell, and Neto}}]{TTCC07}
\bibinfo{author}{\bibfnamefont{K.-M.} \bibnamefont{Tam}},
  \bibinfo{author}{\bibnamefont{S.-W.Tsai}},
  \bibinfo{author}{\bibfnamefont{D.~K.} \bibnamefont{Campbell}},
  \bibnamefont{and} \bibinfo{author}{\bibfnamefont{A.~H.~C.}
  \bibnamefont{Neto}}, \bibinfo{journal}{Phys. Rev. B}
  \textbf{\bibinfo{volume}{75}}, \bibinfo{pages}{161103}
  (\bibinfo{year}{2007}).

\bibitem[{\citenamefont{Bakrim and Bourbonnais}(2007)}]{BB07}
\bibinfo{author}{\bibfnamefont{H.}~\bibnamefont{Bakrim}} \bibnamefont{and}
  \bibinfo{author}{\bibfnamefont{C.}~\bibnamefont{Bourbonnais}},
  \bibinfo{journal}{Phys. Rev. B} \textbf{\bibinfo{volume}{76}},
  \bibinfo{pages}{195115} (\bibinfo{year}{2007}).

\bibitem[{\citenamefont{Honerkamp}(2008)}]{H08}
\bibinfo{author}{\bibfnamefont{C.}~\bibnamefont{Honerkamp}},
  \bibinfo{journal}{Phys. Rev. Lett.} \textbf{\bibinfo{volume}{100}},
  \bibinfo{eid}{146404} (\bibinfo{year}{2008}).

\bibitem[{\citenamefont{Gonzalez}(2008)}]{G08}
\bibinfo{author}{\bibfnamefont{J.}~\bibnamefont{Gonzalez}}
  (\bibinfo{year}{2008}), \eprint{arXiv:0807.3914}.

\bibitem[{\citenamefont{Shankar}(1994)}]{S94}
\bibinfo{author}{\bibfnamefont{R.}~\bibnamefont{Shankar}},
  \bibinfo{journal}{Rev. Mod. Phys.} \textbf{\bibinfo{volume}{66}},
  \bibinfo{pages}{129} (\bibinfo{year}{1994}).

\bibitem[{\citenamefont{Zanchi and Schulz}(2000)}]{ZS00}
\bibinfo{author}{\bibfnamefont{D.}~\bibnamefont{Zanchi}} \bibnamefont{and}
  \bibinfo{author}{\bibfnamefont{H.~J.} \bibnamefont{Schulz}},
  \bibinfo{journal}{Phys. Rev. B} \textbf{\bibinfo{volume}{61}},
  \bibinfo{pages}{13609} (\bibinfo{year}{2000}).

\bibitem[{\citenamefont{Guinea et~al.}(2004)\citenamefont{Guinea, Markiewicz,
  and Vozmediano}}]{GMV04}
\bibinfo{author}{\bibfnamefont{F.}~\bibnamefont{Guinea}},
  \bibinfo{author}{\bibfnamefont{R.~S.} \bibnamefont{Markiewicz}},
  \bibnamefont{and} \bibinfo{author}{\bibfnamefont{M.~A.~H.}
  \bibnamefont{Vozmediano}}, \bibinfo{journal}{Phys.Rev.B}
  \textbf{\bibinfo{volume}{69}}, \bibinfo{pages}{054509}
  (\bibinfo{year}{2004}).

\bibitem[{\citenamefont{Nishine et~al.}(2008)\citenamefont{Nishine, Tsuchiizu,
  and Suzumura}}]{NTS08}
\bibinfo{author}{\bibfnamefont{T.}~\bibnamefont{Nishine}},
  \bibinfo{author}{\bibfnamefont{M.}~\bibnamefont{Tsuchiizu}},
  \bibnamefont{and} \bibinfo{author}{\bibfnamefont{Y.}~\bibnamefont{Suzumura}},
  \bibinfo{journal}{J. Phys.: Conf. Ser.} \textbf{\bibinfo{volume}{132}},
  \bibinfo{pages}{012020} (\bibinfo{year}{2008}).

\bibitem[{\citenamefont{Rold\'{a}n et~al.}(2006)\citenamefont{Rold\'{a}n,
  L\'{o}pez-Sancho, Guinea, and Tsai}}]{R06b}
\bibinfo{author}{\bibfnamefont{R.}~\bibnamefont{Rold\'{a}n}},
  \bibinfo{author}{\bibfnamefont{M.~P.} \bibnamefont{L\'{o}pez-Sancho}},
  \bibinfo{author}{\bibfnamefont{F.}~\bibnamefont{Guinea}}, \bibnamefont{and}
  \bibinfo{author}{\bibfnamefont{S.-W.} \bibnamefont{Tsai}},
  \bibinfo{journal}{\prb} \textbf{\bibinfo{volume}{74}},
  \bibinfo{pages}{235109} (\bibinfo{year}{2006}).

\bibitem[{\citenamefont{Choy et~al.}(2006)\citenamefont{Choy, Cohen, and
  Louie}}]{CCL06}
\bibinfo{author}{\bibfnamefont{H.}~\bibnamefont{Choy}},
  \bibinfo{author}{\bibfnamefont{M.}~\bibnamefont{Cohen}}, \bibnamefont{and}
  \bibinfo{author}{\bibfnamefont{S.}~\bibnamefont{Louie}},
  \bibinfo{journal}{Phys. Rev. B} \textbf{\bibinfo{volume}{73}},
  \bibinfo{pages}{104520} (\bibinfo{year}{2006}).

\bibitem[{\citenamefont{Jiang et~al.}(1996)\citenamefont{Jiang, Schachinger,
  Carbotte, Basov, and Timusk}}]{JSCBT96}
\bibinfo{author}{\bibfnamefont{C.}~\bibnamefont{Jiang}},
  \bibinfo{author}{\bibfnamefont{E.}~\bibnamefont{Schachinger}},
  \bibinfo{author}{\bibfnamefont{J.}~\bibnamefont{Carbotte}},
  \bibinfo{author}{\bibfnamefont{D.}~\bibnamefont{Basov}}, \bibnamefont{and}
  \bibinfo{author}{\bibfnamefont{T.}~\bibnamefont{Timusk}},
  \bibinfo{journal}{Phys. Rev. B} \textbf{\bibinfo{volume}{54}},
  \bibinfo{pages}{1264} (\bibinfo{year}{1996}).

\bibitem[{\citenamefont{Schachinger and Carbotte}(2008)}]{SC08}
\bibinfo{author}{\bibfnamefont{E.}~\bibnamefont{Schachinger}} \bibnamefont{and}
  \bibinfo{author}{\bibfnamefont{J.}~\bibnamefont{Carbotte}},
  \bibinfo{journal}{Phys. Rev. B} \textbf{\bibinfo{volume}{77}},
  \bibinfo{pages}{094524} (\bibinfo{year}{2008}).

\bibitem[{\citenamefont{Leavens and Carbotte}(1971)}]{LC71}
\bibinfo{author}{\bibfnamefont{C.~R.} \bibnamefont{Leavens}} \bibnamefont{and}
  \bibinfo{author}{\bibfnamefont{J.~P.} \bibnamefont{Carbotte}},
  \bibinfo{journal}{Can. J. Phys.} \textbf{\bibinfo{volume}{49}},
  \bibinfo{pages}{724} (\bibinfo{year}{1971}).

\bibitem[{\citenamefont{Allen and Dynes}(1975)}]{AD75}
\bibinfo{author}{\bibfnamefont{P.~B.} \bibnamefont{Allen}} \bibnamefont{and}
  \bibinfo{author}{\bibfnamefont{R.~C.} \bibnamefont{Dynes}},
  \bibinfo{journal}{Phys. Rev. B} \textbf{\bibinfo{volume}{12}},
  \bibinfo{pages}{905} (\bibinfo{year}{1975}).

\bibitem[{\citenamefont{Daams and Carbotte}(1981)}]{DC81}
\bibinfo{author}{\bibfnamefont{J.~M.} \bibnamefont{Daams}} \bibnamefont{and}
  \bibinfo{author}{\bibfnamefont{J.~P.} \bibnamefont{Carbotte}},
  \bibinfo{journal}{J. of Low Temp. Phys.} \textbf{\bibinfo{volume}{43}},
  \bibinfo{pages}{263} (\bibinfo{year}{1981}).

\bibitem[{\citenamefont{Tsai et~al.}(2006)\citenamefont{Tsai, {Castro Neto},
  Shankar, and Campbell}}]{Shan05b}
\bibinfo{author}{\bibfnamefont{S.-W.} \bibnamefont{Tsai}},
  \bibinfo{author}{\bibfnamefont{A.~H.} \bibnamefont{{Castro Neto}}},
  \bibinfo{author}{\bibfnamefont{R.}~\bibnamefont{Shankar}}, \bibnamefont{and}
  \bibinfo{author}{\bibfnamefont{D.~K.} \bibnamefont{Campbell}},
  \bibinfo{journal}{Phil. Mag.} \textbf{\bibinfo{volume}{86}},
  \bibinfo{pages}{2631} (\bibinfo{year}{2006}).

\bibitem[{\citenamefont{Polchinski}(1994)}]{P94}
\bibinfo{author}{\bibfnamefont{J.}~\bibnamefont{Polchinski}},
  \bibinfo{journal}{Nucl. Phys. B} \textbf{\bibinfo{volume}{422}},
  \bibinfo{pages}{617} (\bibinfo{year}{1994}).

\bibitem[{\citenamefont{Carbone et~al.}(2008)\citenamefont{Carbone, Yang, and
  Zwail}}]{CYGZ08}
\bibinfo{author}{\bibfnamefont{F.}~\bibnamefont{Carbone}},
  \bibinfo{author}{\bibfnamefont{D.~S.} \bibnamefont{Yang}}, \bibnamefont{and}
  \bibinfo{author}{\bibfnamefont{A.~H.} \bibnamefont{Zwail}},
  \bibinfo{journal}{Proc. Nat. Acd. Sci. USA} \textbf{\bibinfo{volume}{105}},
  \bibinfo{pages}{20161} (\bibinfo{year}{2008}).

\bibitem[{\citenamefont{Maksimov et~al.}(2008)\citenamefont{Maksimov, Kulc, and
  Dolgov}}]{MKD08}
\bibinfo{author}{\bibfnamefont{E.~G.} \bibnamefont{Maksimov}},
  \bibinfo{author}{\bibfnamefont{M.~L.} \bibnamefont{Kulc}}, \bibnamefont{and}
  \bibinfo{author}{\bibfnamefont{O.~V.} \bibnamefont{Dolgov}}
  (\bibinfo{year}{2008}), \eprint{arXiv:0810.3789}.

\end{thebibliography}

\end{document}